\documentclass[useAMS,usenatbib]{mn2e}

\usepackage{times}
\usepackage{graphicx}
\usepackage{natbib}
\usepackage{txfonts}
\usepackage{subfigure}
\usepackage{footnote}

\title[Revised distance to the 1806--20 cluster]{A downward revision to the
distance of the 1806--20 cluster and associated magnetar from Gemini 
near-Infrared spectroscopy}

\author[J. L. Bibby et al.]{J. L. 
Bibby,$^{1}$, P. A. Crowther,$^{1}$ J. P. Furness$^{1}$ and J. S. 
Clark$^{2}$. \\
$^{1}$University of Sheffield, Department of Physics \&
              Astronomy, Hicks Building, Hounsfield Rd, Sheffield, S3
              7RH, E-mail: j.bibby@sheffield.ac.uk \\
$^{2}$Department of Physics \& Astronomy, The Open University, Walton Hall, Milton Keynes, MK7 6AA }

\begin{document}

\date{January 2008}

\pagerange{\pageref{firstpage}--\pageref{lastpage}} \pubyear{2008}

\maketitle

\label{firstpage}

\begin{abstract}{We present $H$- and $K$-band spectroscopy of OB and 
Wolf-Rayet (WR) members of the Milky Way cluster 1806--20
(G10.0--0.3), to obtain
a revised cluster distance of relevance to the 2004 giant flare from
the SGR 1806--20 magnetar. From GNIRS spectroscopy 
obtained with Gemini South, four candidate OB
stars are confirmed as late O/early B supergiants, while we support previous
mid WN and late WC classifications for two WR stars.  Based upon an
absolute $K_{s}$-band magnitude calibration for B supergiants and WR
stars, and near-IR photometry from NIRI at Gemini 
North plus archival VLT/ISAAC datasets,
we obtain a cluster distance modulus of 14.7$\pm$0.35 mag. The
known stellar content of the 1806--20 cluster suggests an age of 3--5 Myr, 
from which theoretical isochrone fits infer a distance 
modulus of 14.7$\pm$0.7 mag. Together, our results favour a distance
modulus of 14.7$\pm$0.4 mag (8.7$^{+1.8}_{-1.5}$ kpc) to the 1806--20
cluster, which is significantly lower than the nominal 15 kpc distance
to the magnetar. For our preferred distance, the peak luminosity of
the December 2004 giant flare is reduced by a factor of three to 7
$\times$ 10$^{46}$ ergs$^{-1}$, such that the contamination of BATSE
short gamma ray bursts (GRB's) from giant flares of extragalactic
magnetars is reduced to a few percent. We infer a magnetar progenitor
mass of $\sim48^{+20}_{-8}$ M$_{\odot}$, in close agreement with that
obtained recently for the magnetar in Westerlund~1.}
\end{abstract}

\begin{keywords}
stars: Wolf-Rayet - stars: early-type - Open clusters
and associations: individual: 1806--20 - pulsars: individual: SGR 
1806--20 - Galaxy: kinematics and dynamics 
\end{keywords}

\section{Introduction}

Magnetars are highly magnetized neutron stars, representing a small
subset of slowly rotating pulsars undergoing rapid spin down, and are
observationally associated with Anomalous X-ray Pulsars (AXP's) and
Soft Gamma Repeaters (SGR's). To date only four examples of SGR's are
known, characterised by multiple, soft gamma-ray bursts, typically
10$^{41}$ erg s$^{-1}$ in peak luminosity, plus rare giant flares of
10$^{45}$ erg s$^{-1}$ peak luminosity.  One such giant flare, from
SGR 1806--20 \citep{Kouveliotou1998}, was detected on 27 December 2004
by many satellites including the Burst Alert Telescope (BAT) on Swift
\citep{Palmer2005}, and in January 2005 the radio afterglow was
detected by \citet{Cameron2005} through VLA observations.

It is believed that massive stars are the progenitors of magnetars, on
the basis that several magnetars are associated with young massive
clusters, including SGR 1806--20 which is apparently within a cluster at
G10.0-0.3, forming part of the W31 complex. However, a
distance of 12--15 kpc has been proposed for the 1806--20 cluster
\citep{Eikenberry2004, Figer2004}, yet direct H\,I measurements for the
magnetar suggest 6--10 kpc \citep{Cameron2005}, though the latter has
been questioned by \citet{McClure-Griffiths&Gaensler2005}.

Adopting a distance of 15 kpc suggests SGR 1806--20 had a peak luminosity
of 2 $\times 10^{47}$ erg s$^{-1}$ during the December 2004 giant flare, 
meaning magnetars at distances of up to 30 Mpc could be mistaken for short
gamma-ray bursts (GRB's). For this distance, as much
as 40\% of all short GRB's identified by the Burst And Transient
Source Experiment (BATSE) might be giant flares\citep{Hurley2005},
falling to just 3\% if a distance of $\sim$6 kpc were adopted for 
SGR 1806--20.

%_____________________________________________________________________________
\begin{figure*}
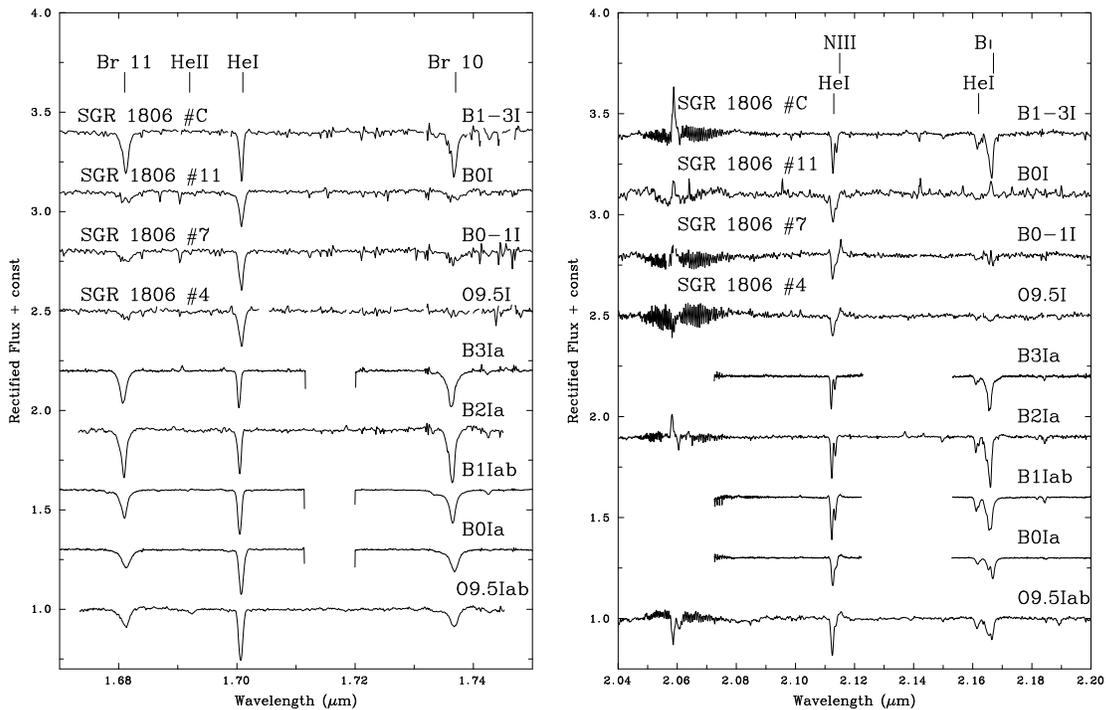

\begin{center}
\mbox{
  \subfigure{\includegraphics[bb=45 110 505 720, 
width=0.4\textwidth]{sgr1806-OB-H-2.ps}}
  \hspace*{0.2cm} 
  \subfigure{\includegraphics[bb=45 110 505 720, 
width=0.4\textwidth]{sgr1806-OB-K-2.ps}} 
}
\caption{Spectral comparison between OB stars in the 1806--20 
cluster and template 
spectra from \citet{Hanson2005} in the $H$- Band (left) and $K$- Band 
(right). 
The structure around $\lambda$\=2.06$\mu$m 
is due to the imperfect telluric correction. }
\label{sgr1806-OB}
\end{center}
\end{figure*}

%____________________________________________________________

The cluster contains young massive stars including OB stars, WR stars
and a Luminous Blue Variable (LBV) \citep{vanKerkwijk, Figer2004, Eikenberry2004}.
If the magnetar is physically associated with this cluster and lies at
15 kpc then a high mass for the LBV, and hence the magnetar
progenitor, of $>$ 133 M$_{\odot}$ is indicated
\citep{Eikenberry2004}. However this is much larger than the 40--55
M$_{\odot}$ progenitor of the AXP found in Westerlund 1, whose stellar
content is reminiscent of the SGR 1806--20 cluster
\citep{Muno2006,Clark2008}.

In this letter, spectra of several massive stars  within
the cluster are presented in Section 2. In Section 3, spectral types are determined
together with estimates of absolute magnitude for each star. This
allows the cluster distance and age to be obtained in Section 4 and
compared to that of the magnetar. Conclusions are drawn in Section 5.

%__________________________________________________________________
\section{Observations}
Spectroscopic data of 1806--20 cluster members
were obtained from the Gemini Near-Infrared
Spectrograph (GNIRS) instrument at Gemini South. New near-IR
photometry was provided by the NIRI instrument at Gemini 
North\footnotemark[1],
supplemented by Very Large Telescope (VLT) Infrared Spectrometer And Array 
Camera (ISAAC) and New Technology Telescope (NTT) 
Son of ISAAC (SofI) archival observations.

\footnotetext[1]{Following severe damage to GNIRS in April 2007,
spectroscopy of additional 1806--20 cluster members were obtained with 
NIRI at Gemini North, but these did not 
improve on previously published work due to a misaligned
slit and low spectral resolution..}

%______________________________________________________________
\subsection{Photometry}

NIRI $H$- band acquisition images of the 1806--20 cluster were
obtained on July 13-14 2007 during conditions with an image quality of
$\sim$0.35 arcsec (3 pix) FWHM. 2MASS $H$- band photometry of isolated,
bright sources 2, 6, 11, and 12 from \citet{Figer2004} provided a
zero point, from which errors of $\pm$0.03 mag were obtained. In addition
$K_{s}$- band archival VLT ISAAC images from March 17 2005 (program
274.D.5048) obtained with an image quality of 0.4 arcsec (2.4 pixel
FWHM), plus 2MASS $K_{s}$- band photometry of the same bright sources
as used with NIRI provided zero points and errors of $\pm$0.02 mag. In
the case of star B from \citet{Eikenberry2004} the ISAAC images were
saturated. For this star we resorted to archival NTT SofI $K_{s}$ band
images from July 31 2003 (program 271.D.5041) for which an image
quality of FWHM $\sim$1.1 arsec (8 pix) was measured, again with zero
points from 2MASS and errors of $\pm$0.1 mag.

%______________________________________________________________

%__________________________________________________________________

\subsection{Spectroscopy}

We used the GNIRS instrument on the 8 m Gemini South telescope at
Cerro Pachon, Chile on April 8--9 2007 to observe a number of massive
stars associated with cluster 1806--20 during $K-$-band 
seeing conditions of FWHM $\sim$0.5 arcsec. Spectroscopic data were
obtained in the $H$- band over a wavelength range 1.625--1.775 $\mu$m and
in the $K$- band for 2.025--2.225 $\mu$m. The slit width was 3
pixels (0.45 arcseconds), providing a
resolution R $\sim$3700. Individual exposures of 270s ($H$- band) and
80s ($K$- band) were combined producing total exposures of
640--960 seconds ($K$- band) and 2160--3240 seconds ($H$- band).

The telescope was nodded along three slit positions of differing
position angle, allowing observations of star C from
\citet{Eikenberry2004} plus stars 1, 2, 4, 7, and 11 from
\citet{Figer2005}, which allowed pairs of images for sky subtraction.
Internal Argon lamp images were obtained immediately before or after
each slit position, providing an accurate wavelength calibration.
Further spectra were obtained for the telluric standard star HIP89384
(B6\,V) with an identical setup, except reduced exposure times of 30
seconds in both the $H$- and $K$- band, for which sky lines were used
to provide a wavelength calibration. Calibration flat fields and dark
frames were also obtained.

 \begin{figure}
   \centering
   \includegraphics[bb=35 105 480 
   725,width=0.3\textwidth,angle=-90]{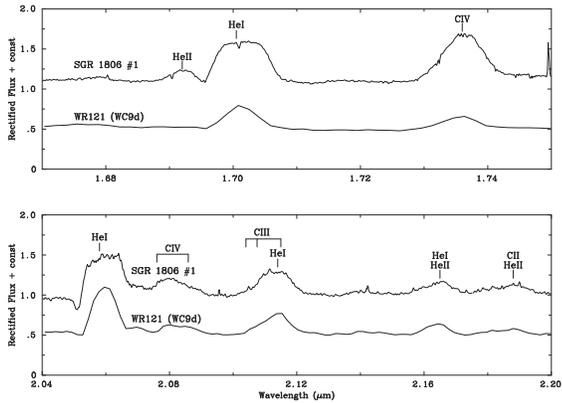}
     \caption{Spectral comparison of \#1 from \citet{Figer2005} in the
    1806--20 cluster with the Galactic dusty WC9 star WR121 in the
     $H$ (upper panel) and $K$ (lower panel) band.}
                   \label{sgr1806-1.ps}
   \end{figure}

Initial data reduction (sky subtraction, flat fielding and wavelength
calibration) was done using the Image Reduction and Analysis Facility
(IRAF)\footnotemark[2] then spectra for each object observed were
extracted. STARLINK packages were then used to telluric correct and
normalise the spectra. Stellar Brackett lines from the telluric
standard star were fit using empirical profiles from high spectral
resolution observations of late B/early A dwarfs of
\citet{Hanson2005}.

\footnotetext[2]{IRAF is distributed by the National Optical Astronomy Observatories,
    which are operated by the Association of Universities for Research
    in Astronomy, Inc., under cooperative agreement with the National
    Science Foundation.}

%_____________________________________________________________

\section{Spectroscopy of 1806--20 Cluster Members}

We supplemented GNIRS observations of stars 1, 2, 4, 7, and 11 from
\citet{Figer2005} with their observations of star 3 plus GNIRS
spectroscopy of Star C from \citet{Eikenberry2004} with their
observations of stars B and D. After reduction and calibration, the
observed spectra were compared to template stars from the spectral atlases of
\citet{Hanson2005} for OB stars and \citet{CrowtherandSmith1996} for
WR stars.

%__________________________________________________________________

 \begin{figure}
   \centering
   \includegraphics[bb=35 105 480 
   725,width=0.3\textwidth,angle=-90]{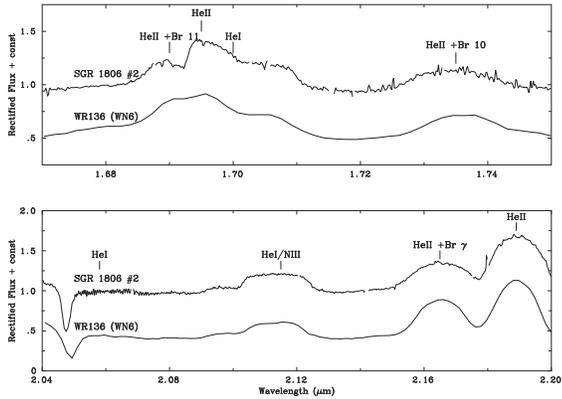}
     \caption{Spectral comparison of \#2 from \citet{Figer2005} in the 
     1806--20 cluster with WR136 (WN6b) in the $H$ (upper panel) and
     $K$ (lower panel) band.}
                   \label{sgr1806-2.ps}
   \end{figure}

%___________________________________________________________________
%______________________________________________________________

\subsection{OB Stars}

In Figure~\ref{sgr1806-OB}, GNIRS spectra of cluster OB stars are
compared to template supergiants from \citet{Hanson2005}. In the $H$-
band, stars C and 7 clearly shows
He\,I~1.700$\mu$m absorption but the
He\,II~1.692$\mu$m line is not present, indicating that they
are early B stars. Star C shows strong Br 10
absorption supporting a classification of B1--3\,I. Similarly, star 7 shows a
relatively weak Br 10 line, indicating B0--1\,I. $K$- band observations 
support these classifications since the presence of the
N\,III~2.115$\mu$m emission in star 7 is common to B0--1
supergiants, where N\,III~2.115$\mu$m absorption in star C is
shared with B1--3 supergiants. Later subtypes are excluded from
the observed strength of He\,I features, and again no evidence for
He\,II~2.189$\mu$m is seen.

\begin{table}
\caption{Adopted absolute $K_{s}$- band magnitude calibration for early B
supergiants based on absolute visual magnitudes from Conti, Crowther 
\& Leitherer (2008), intrinsic near IR colours and bolometric corrections (BC$_{K_{s}}$)
from Crowther et al.  (2006a) }
\begin{tabular}{cccccc}
\hline\hline
Sp Type&$M_{V}$         & ($V-K_{s}$)$_{0}$ & ($H-K_{s}$)$_{0}$ & 
$M_{K_{s}}$           & $BC_{K_{s}}$\\
\hline
O9.5\,I& --6.4 $\pm$ 0.5 & --0.71            & --0.08            & --5.7 $\pm$ 0.5       & --3.4 \\
B0\,I  & --6.6 $\pm$ 0.5 & --0.75            & --0.08            & --5.85 $\pm$ 0.5      & --3.3 \\
B1\,I  & --6.9 $\pm$ 0.5 & --0.62            & --0.08            & --6.3 $\pm$ 0.5     & --2.65\\
B2\,I  & --7.1 $\pm$ 0.5 & --0.53            & --0.08            & --6.6 $\pm$ 0.5     & --2.1 \\
B3\,I  & --7.1 $\pm$ 0.5 & --0.36            & --0.05            & --6.7 $\pm$ 0.5     & --1.7\\
\hline

\end{tabular}
\label{table1}
\end{table}

Star 11 shows relatively weak Br 11~1.681$\mu$m and Br
10~1.736$\mu$m absorption in the $H$- band,
indicating an B0\,I spectral type, whilst its $K$- band spectra
is ambiguous due to the lack of Br $\gamma$, as 
noted by \citet{Figer2005}. Weak He\,II~2.189$\mu$m absorption 
is present in star 4, supporting an O9.5\,I
spectral classification.

In all these cases OB supergiant
classifications proposed by \citet{Eikenberry2004} and
\citet{Figer2005} are supported for which we are able to provide more
refined subtypes, albeit unable to distinguish between Ia and Ib luminosity classes.

%____________________________________________________________________

\begin{table*}
\begin{center}
\caption{Spectral classifications and photometry of 1806--20
cluster members together with interstellar extinctions $A_{K_{s}}$,
adopted absolute magnitudes $M_{K_{s}}$ and resulting distance moduli
($DM$).}
\begin{tabular}{ccccccccc}
\hline\hline
Star & Spect.    & $K_{s}$ & $H-K_s$ & $(H-K_s)_{0}$ & 
$E_{\rm H-K_{s}}$  & $A_{K_{s}}$ & $M_{K_{s}}$ & $DM$\\
     & Type     & mag      & mag     & mag & mag & mag & mag & mag \\
\hline
\#1 & WC9d        & 11.60   & 2.16  & +1.10   & 1.06                                                               \\
\#2 & WN6b        & 12.16   & 1.89  & +0.27   & 1.62  & 2.95 $\pm$ 0.5   & --4.77 $\pm$ 0.7      & 13.98 $\pm$ 0.86\\
\#3 & WN7         & 12.58   & 1.67  & +0.11   & 1.56  & 2.84 $\pm$ 0.48    & --5.92 $\pm$ 0.8    & 15.93 $\pm$ 0.93\\
\#4 & O9.5\,I     & 11.92   & 1.55  & --0.09 & 1.64  & 2.98 $\pm$ 0.5     & --5.7 $\pm$ 0.5     & 14.64 $\pm$ 0.71\\
\#7 & B0--B1\,I   & 11.87   & 1.55  & --0.08 & 1.63  & 2.97 $\pm$ 0.49    & --6.0 $\pm$ 0.8     & 14.90 $\pm$ 0.94\\
\#11& B0\,I       & 11.90   & 1.63  & --0.08 & 1.71  & 3.11 $\pm$ 0.52    & --5.85 $\pm$ 0.5    & 14.64 $\pm$ 0.72\\
  B & WC9d        & 10.40   & 3.03  & +1.10   & 1.93                                                               \\
  C & B1--B3\,I   & 10.96   & 1.81  & --0.08 & 1.89  & 3.43 $\pm$ 0.57    & --6.50 $\pm$ 0.8    & 13.88 $\pm$ 0.98\\
  D & OB\,I         & 11.06   & 1.69  & --0.09 & 1.78                                                               
\\
\hline
\textbf{Average}& &         &       &       &       & 3.00 $\pm$ 0.3     &                     & 14.69 $\pm$ 0.35\\
\hline

\end{tabular}
\label{table2}
\end{center}
\end{table*}
%______________________________________________________________________________

%_____________________________________________________________

\subsection{WR Stars}

Star 1 from \citet{Figer2005} has previously been identified as a
WC8 star. However inspection of GNIRS spectroscopy in
Figure~\ref{sgr1806-1.ps} reveals that the relative strength of the
C\,IV~1.736$\mu$m to the He\,I~1.700$\mu$m emission is too low for a WC8
star and is more typical of dusty WC9 stars such as WR121. Moreover,
the C\,IV~2.076$\mu$m/C\,III~2.110$\mu$m ratio is consistent with other
observations of dusty WC9 stars by Crowther et al. 
(2006a)\nocite{Crowther2006}. Warm 
circumstellar dust contributes to their 
near IR flux, resulting in
emission lines appearing relatively weak with respect to non-dusty WC9
stars, for example WR88 from \citet{EenensWilliams&Wade1991}, from
which a WC9d spectral type results for star 1.

The GNIRS spectra for star 2 presented in Figure~\ref{sgr1806-2.ps}
produces a classification of a broad-lined WN6 star (WN6b), in
agreement with \citet{Figer2005}. This is evident from the strong
He\,II~2.189$\mu$m emission compared to the weaker
He\,I/N\,III~2.115$\mu$m and Br $\gamma$ emission. A later WN
classification would require a higher Br $\gamma$ to He\,II 2.189$\mu$m ratio
(Crowther et al. 2006a)\nocite{Crowther2006}.
%_____________________________________________________________________________

\section{Distance to 1806--20 cluster}
We are now in a position to determine the distance to the 1806--20
cluster from both spectroscopic methods, using absolute magnitude
versus spectral type calibrations for OB and WR stars and from fitting
isochrones to the position of OB supergiants in the Hertzsprung-Russell 
(HR) diagram.
%_____________________________________________________________
\subsection{Absolute Magnitude Calibration}

The near IR absolute magnitude calibration for O stars
\citep{Martins&Plez2006} does not extend to early B
supergiants. Consequently absolute \textit{visual} magnitude calibration
of B supergiants from Conti et al. (2008)\nocite{Conti2008} are used 
together with
synthetic near-infrared intrinsic colours of early B supergiants from
Crowther, Lennon \& Walborn (2006b)\nocite{CrowtherLennonWalborn2006} to 
obtain a $K_{s}$-band absolute
magnitude calibration, as shown in Table~\ref{table1}. The empirical spread
in our OB supergiant calibration is estimated to be $\pm$0.5 mag.
For comparison, at O9.5\,I, which is 
common to both calibrations, \citet{Martins&Plez2006} obtain M$_{K}$ =
--5.52 mag, $(H-K)$ = --0.10 mag and BC$_{K}$ = --3.66 mag. 

For WR stars, absolute magnitudes from Crowther et al. 
(2006a)\nocite{Crowther2006} were used
for calibration, with a typical spread of $\pm$0.7 mag. Intrinsic
colours are drawn from the same sources with typical uncertainties of
$\pm$0.05 mag. Dusty WC stars span a much wider range in both absolute
magnitude (M$_{K}$ = --8.5$\pm$1.5 mag) and intrinsic colour [($H-K$)$_{o}$ 
= 1.1$\pm$0.6] at near IR wavelengths than normal WC
stars, preventing their use as reliable distance indicators (Crowther
et al. 2006a\nocite{Crowther2006}).

%___________________________________________________________

\subsection{Spectroscopic Distance}

Observed colours of OB and WR stars in the 1806--20 cluster provide
a direct measurement of interstellar extinction from $ A_{K_{s}} =
1.82^{+0.30}_{-0.23} E_{\rm H-K_{s}}$ 
\citep{Indebetouw2005}. These
are presented in Table~\ref{table2}, from which a mean
$A_{K_{s}}$$\sim$3.0$\pm$0.3  mag is obtained, together with distance 
moduli for each star, based upon absolute magnitude calibrations, from
which dusty WC stars were excluded, as explained above. A mean distance
modulus of 14.69$\pm$0.35 mag results from this spectroscopic study,
corresponding to 8.7$^{+1.5}_{-1.3}$ kpc.

%____________________________________________________________________
\subsection{Distance from Isochrone Fitting}

Alternatively, theoretical isochrones may be used to estimate the
distance to the 1806--20 cluster if limits upon the cluster age are
available. As noted above, this cluster hosts OB stars, an LBV,
together with helium burning WN and WC stars, a stellar content in
common with other massive star clusters within the inner Milky
Way. These include Westerlund 1 and
the Quintuplet cluster, for which ages of 4-5 Myr and 4$\pm$1 Myr
have been inferred by Crowther et al. (2006a)\nocite{Crowther2006} and 
\citet{Figer1999}, respectively.

OB supergiants \#4 (O9.5\,I) and 11 (B0\,I) have the best determined
spectral types, from which stellar temperatures (29kK and 27.5kK) and
luminosities can be derived (Crowther et
al. 2006b\nocite{CrowtherLennonWalborn2006}), the latter obtained from
$K_{s}$-band bolometric corrections (BC$_{K_{s}}$, see
Table~\ref{table1}). For a variety of different adopted distances,
these two stars provide cluster ages using isochrones from
\citet{Lejeune&Schaerer2001}, which are based upon high mass-loss
rate, solar metallicity evolutionary models from
\citet{Meynet1994}. Table 3 presents ages, OB supergiant and (minimum)
magnetar initial masses for distance moduli in the range 14.0 (6.3
kpc) to 15.9 (15 kpc).  Large distances would require these stars to be
hypergiants, characterised by an emission line spectrum, for
which there is no spectroscopic evidence. From above 
a distance modulus of 14.7$\pm$0.7 mag (6.3--12
kpc) is obtained, together with a magnetar progenitor mass of 35--100
M$_{\odot}$.

%____________________________________________________________________

\subsection{Comparison with Previous Work}

Together these two methods of estimating the distance to the 1806--20
cluster suggest a distance modulus of 14.7$\pm$0.4 mag
(8.7$^{+1.8}_{-1.5}$ kpc). This is substantially lower the nominal
$\sim$15 kpc kinematic distance to the magnetar obtained by
\citet{Corbel1997} from CO observations which was supported by
\citet{Eikenberry2004} from the nebular Br $\gamma$ emission (Local
Standard of Rest) velocity for LBV 1806--20. In Figure 4 the cluster
distance from this work is compared to \citet{Eikenberry2004} and
\citet{Figer2004}, adapted to a Galactic Centre distance of 8 kpc
\citep{Reid1993} and the \citet{Brand&Blitz1993} rotation model. In
addition, magnetar distances from \citet{Corbel1997},
\citet{Cameron2005} and \citet{McClure-Griffiths&Gaensler2005} are
included, also updated for a Galactic Centre distance of 8 kpc.

Our distance reconciles the previously inconsistent cluster and
magnetar distances. In addition, OB supergiant stellar He\,I
1.700$\mu$m and Br$\gamma$ line profiles from GNIRS reveal
V$_{\rm{LSR}}\sim$ 130 km\,s$^{-1}$, indicating kinematic distances of
either 6.8 kpc or 8.9 kpc, in good agreement with both the
spectroscopic parallax and isochrone fitting methods (previous studies
implied V$_{\rm{LSR}}$=10--35 km\,s$^{-1}$, from \citet{Corbel1997}
and \citet{Figer2004})

%____________________________________________________________________

\begin{table}
\begin{center}
\caption{Ages and progenitor masses ($M_{\rm init}^{\rm OB}$) of \#4 
(O9.5\,I) and 
\#11 (B0\,I)
from \citet{Lejeune&Schaerer2001} isochrones using \citet{Meynet1994}
evolutionary tracks for a variety of distance moduli ($DM$) plus inferred 
minimum magnetar progenitor masses ($M_{\rm init}^{\rm SGR}$).}
\begin{tabular}{c
@{\hspace{3mm}}c
@{\hspace{3mm}}c
@{\hspace{3mm}}c
@{\hspace{3mm}}c
@{\hspace{3mm}}c
@{\hspace{3mm}}c
@{\hspace{3mm}}c
@{\hspace{3mm}}c}
\hline\hline
$DM$  & $d$ & Star & $M_{K_{s}}$ & $M_{\rm Bol}$  & Log $T$ & Age & 
$M_{\rm init}^{\rm OB}$ & 
$M_{\rm init}^{\rm SGR}$ \\
(mag)    & (kpc)    &      & (mag)       & (mag)      &  (K)  &(Myr)& 
(M$_{\odot}$)&  (M$_{\odot}$) \\
\hline
14.0 & 6.3     & \#4  & --5.1       & --8.5      & 4.46    & 5 & 30        
& 35 \\
     &         & \#11 & --5.2       & --8.5      & 4.44    &   & 30
& 35 \\
14.3 & 7.2     & \#4  & --5.4       & --8.8      & 4.46    & 4.6&33        
& 40 \\   
     &         & \#11 & --5.5       & --8.8      & 4.44    &   & 33        
& 40 \\
14.7 & 8.7     & \#4  & --5.8       & --9.2      & 4.46    & 4 & 40        
& 48 \\
     &         & \#11 & --5.9       & --9.2      & 4.44    &   & 40        
& 48 \\
15.1 & 10.5    & \#4  & --6.2       & --9.6      & 4.46    & 3.4&49        
& 69 \\ 
     &         & \#11 & --6.2       & --9.6      & 4.44    &   & 49        
& 69 \\
15.4 & 12      & \#4  & --6.5       & --9.9      & 4.46    & 3 & 55        
& 100 \\
     &         & \#11 & --6.6       & --9.9      & 4.44    &   & 55        
& 100 \\
15.9 & 15      & \#4  & --7.0       & --10.4     & 4.46    & 2.8 & 80      
& 120 \\
     &         & \#11 & --7.1       & --10.4      & 4.44    &   & 80         
& 120 \\
\hline
\end{tabular}
\label{table3}
\end{center}
\end{table}
%___________________________________________________________________

%______________________________________________________________
\section{Discussion and Conclusions}

A revised distance of $\sim$8.7$^{+1.8}_{-1.5}$ kpc to
the 1806--20 cluster is obtained from this work, reconciling previous cluster and
magnetar distances, from which a magnetar progenitor mass of $\sim
48^{+20}_{-8}$ M$_{\odot}$ is inferred, consistent with estimates for
the Westerlund 1 magnetar (AXP) from \citet{Muno2006} and
\citet{Clark2008}. Similar conclusions were reached by
\citet{Figer2005}. The observed association between magnetars and supernova
remnants suggests that magnetars are young neutron stars ($\leq10^{4}$
yr, \citet{Gaensler2001}), so the actual progenitor mass ought to be
close to the values listed in Table 3.  Following the approach of
\citet{Figer2004}, we estimate a mass of $\sim$36 M$_\odot$ for 
LBV 1806--20 based on
our revised distance, assuming it is a binary with equal mass
components.

This distance represents a major downward revision to the current
adopted magnetar distance of 15 kpc, suggesting a peak luminosity of
$\sim$7$\times$10$^{46}$ ergs$^{-1}$ for the December 2004 giant
flare. \citet{Hurley2005} argue that up to 40$\%$ of all BATSE short
GRB's could be giant flares from magnetars, if one was to adopt a 15
kpc distance to SGR 1806--20 and a frequency of giant flares of one
per 30 yr per Milky Way galaxy. For the revised distance, perhaps only
$\sim$8$\%$ of BATSE short GRB's have an origin in magnetar giant
flares.

 \begin{figure}
   \centering
   \includegraphics[width=0.4\textwidth]{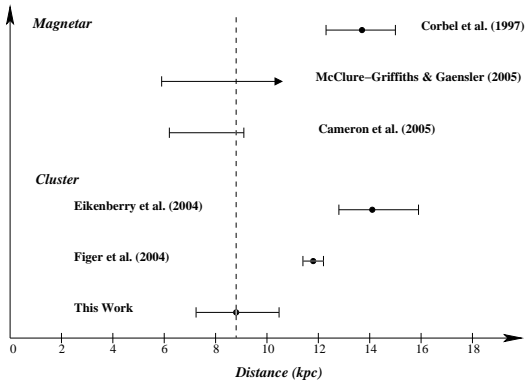}
     \caption{Comparison of present results for the distance to the
    1806--20 cluster with previous cluster and magnetar distance
     estimates, adapted to a Galactic Centre distance of 8 kpc 
     \citep{Reid1993}}
                   \label{sgr1806-reviseddistances.eps}
   \end{figure}

Our preferred distance of
$\sim$8.7$^{+1.8}_{-1.5}$ kpc to the 1806--20 cluster suggests active star
formation at a distance of 1.6$^{+1.4}_{-0.1}$ kpc from the centre of the Milky
Way. It is well known that the amount of molecular hydrogen is greatly
reduced interior to the bar at $\sim$4 kpc \citep{Benjamin2005}
causing a deficiency in HII regions \citep{Russeil2003} in the inner
Milky Way, apart from the Galactic Centre region itself. Apparently
relatively massive clusters can form within this region.
From the current massive star census, a cluster mass in excess of $\sim$3$\times$10$^{3}$
M$_{\odot}$ is estimated from comparison with the Quintuplet
cluster, albeit based upon highly incomplete statistics. Further
spectroscopic studies of the 1806--20 cluster are recommended to
further refine the distance and detailed stellar content.

\section*{Acknowledgements}

Based on observations obtained at the Gemini Observatory, which is
operated by AURA Inc., under a cooperative agreement with the NSF on 
behalf of the Gemini partnership: the NSF (United States),
the STFC (United Kingdom), the
NRC (Canada), CONICYT (Chile), the ARC (Australia), CNPq (Brazil) and 
SECYT (Argentina). This publication makes use of data products from 2MASS, 
which is a joint project of the University of
Massachusetts and the IPAC/CalTech, funded by the NASA and the NSF, and
is based in part on archival observations from the ESO Science Archive 
Facility collected at the La Silla Paranal Observatory. 
JLB and JPF acknowledge financial support from STFC.

\setlength{\bibsep}{0pt}
%\bibliography{SGR1806}

\bibliographystyle{mn2e}
\bibliography{abbrev,SGR1806}

\label{lastpage}

\end{document}